\documentclass[dvips]{article}
\usepackage{icrctc07}

\title{A Search for Prompt Very High Energy Emission from Satellite-detected Gamma-ray Bursts using Milagro}
\shorttitle{Search for VHE Emission from GRBs with Milagro}
\authors{P. M. Saz Parkinson$^{1}$ \& B. L. Dingus$^{2}$ for the Milagro Collaboration.}
\shortauthors{Saz Parkinson \& Dingus}
\afiliations{$^1$ Santa Cruz Institute for Particle Physics, U. of California, 1156 High Street, Santa Cruz, CA 95064, USA\\ $^2$ Group P-23, Los Alamos National Laboratory, P.O. Box 1663, Los Alamos, NM 87545, USA}
\email{pablo@scipp.ucsc.edu; dingus@lanl.gov}

\abstract{Gamma-ray bursts (GRBs) have been detected up to GeV energies and are predicted 
by many models to emit in the very high energy (VHE, $>$ 100 GeV) regime too. Detection 
of such emission would allow us to constrain GRB models. Since its launch, in late 2004, 
the Swift satellite has been locating GRBs at a rate of approximately 100 per year. The 
rapid localization and follow-up in many wavelengths has revealed new and unexpected 
phenomena, such as delayed emission in the form of bright X-ray flares. The Milagro 
gamma-ray observatory is a wide field of view (2 sr) instrument employing a water 
Cherenkov detector to continuously ($>$ 90\% duty cycle) observe the overhead sky in 
the 100 GeV to 100 TeV energy range. Over 100 GRBs are known to have been in the field 
of view of Milagro since January 2000, including 57 since the launch of Swift (through May 2007). 
We discuss the results of the searches for prompt emission from these bursts, as well as for delayed 
emission from the X-ray flares observed in some of the Swift bursts.}

\begin{document}
\maketitle

\section{Introduction}

Almost 40 years after the detection of the first gamma-ray burst (GRB), many questions
remain about these powerful explosions. Progress in the field has accelerated in the 
decade since the detection of the first X-ray afterglow~\cite{costa}. The launch of 
Swift~\cite{gehrels}, in late 2004, has resulted in the rapid and accurate localization 
of $\sim$100 GRBs per year. This has enabled an unprecedented number of multiwavelength follow-up 
observations in every band of the electromagnetic spectrum, from radio to the highest energy gamma 
rays, paving the way for a more complete understanding of these phenomena.

The highest energy emission to be detected conclusively from GRBs was seen by the EGRET 
instrument. EGRET detected several bursts above 100 MeV, with no evidence for a cut-off in the
spectrum~\cite{dingus01}. One burst in particular, GRB 940217, emitted an
18 GeV photon over 90 minutes after the start of the burst~\cite{hurley94}. Another 
burst, GRB 941017, was found to display a second, higher energy, spectral component which 
extended up to at least 200 MeV and decayed more slowly than the lower energy component~\cite{gonzalez03}.
On this evidence alone, the expectations are that the upcoming launch of GLAST, with its much 
higher sensitivity than EGRET, will result in a large number of detections at high energies, leading 
to a better understanding of the GeV properties of GRBs.

At very high energy (VHE, $>$100 GeV), there have been no conclusive detections of GRBs, although there 
have been several tantalizing hints of emission. Milagrito, a 
prototype of Milagro, searched for emission coincident with 54 BATSE bursts and reported 
evidence for emission above 650 GeV from GRB 970417a, at the 3$\sigma$ level~\cite{atkins00a,atkins03}.
The HEGRA group reported evidence at the 3 sigma level for emission above 20 TeV from 
GRB 920925c~\cite{padilla98}. Follow-up observations above 250 GeV by the Whipple 
atmospheric Cherenkov telescope~\cite{connaughton97} failed to find any high energy 
afterglow for 9 bursts studied, though the delay in slewing to observe these bursts 
ranged from 2 minutes to almost an hour. More recent efforts by the MAGIC~\cite{magicGRBlimits}, 
Whipple~\cite{whippleGRBlimits}, and Milagro~\cite{milagro1} telescopes have  
resulted only in upper limits.

There is no shortage of models predicting (or explaining) very high energy emission 
from GRBs (e.g.~\cite{dermer00,pilla98,zhang01,granot03,gupta07}). Many of the proposed
emission mechanisms predict a fluence at TeV energies comparable to that at keV-MeV energies. 
One such mechanism involves the inverse Compton upscattering of lower energy (synchrotron) 
photons by the energetic electrons which emitted them. The strong magnetic fields and large
bulk Lorentz factors present in GRB jets can result in a high energy component peaked at TeV energies. 
The strength of such a component is highly dependent on the environments of the particle 
acceleration and the gamma ray production.

Milagro\cite{atkins00b,atkins01} is a TeV gamma-ray detector, located at an altitude of 
2630 m in northern New Mexico. Milagro uses the water Cherenkov technique to detect 
extensive air-showers produced by VHE gamma rays as they interact 
with the Earth's atmosphere. The Milagro field of view is $\sim$2 sr and duty cycle is $>$90\%. The 
effective area is a function of zenith angle and ranges from $\sim50$ m$^2$ at 100 GeV 
to $\sim10^5$ m$^2$ at 10 TeV. A sparse array of 175 4000-l water tanks, each with a PMT, 
was added in 2002. These ``outriggers,'' extend the physical area of Milagro to 
40000 m$^2$. The combination of large field of view and high duty cycle make Milagro the best 
instrument currently available for conducting a search for prompt VHE emission from GRBs. Milagro 
is also able to operate in ``scaler'' mode, where individual tube rates can be monitored to search for   
GRB emission in the 1--100 GeV energy range coincident in time with known satellite bursts 
(see ~\cite{taylor,cesar}). In addition to conducting a search for emission from satellite-detected GRBs, 
Milagro is capable of performing a blind search for emission at all locations in its field of 
view and many different durations (see ~\cite{vlasios}).

\subsection{The GRB Sample}

Twenty-five satellite-triggered GRBs occurred within the field of view of Milagro 
between January 2000 and December 2001. No significant emission was detected from any of 
these bursts~\cite{atkins05}. In the period from January 2002 to December 2004 (post-BATSE and pre-Swift), 
there were only 11 well-localized GRBs within the Milagro field of view. Since the launch of Swift (through the end of May 2007), 
however, there have been a total of 57 bursts in the field of view of Milagro~\footnote{A good source of information 
on well-localized GRBs is J. Greiner's web page {\tt http://www.mpe.mpg.de/$\sim$jcg/grbgen.html}}, 
many of them with measured redshift. Table~\ref{grb_table} lists the 57 GRBs in this sample, along with some of 
their properties. Due to the absorption of high-energy gamma rays by the extragalactic 
background light (EBL), detections at VHE energies are only expected for redshifts less than $\sim$0.5. The 
degree of gamma-ray extinction from this effect is uncertain, because the amount of EBL is not well known. 
In this work we use the model of \cite{primack05}, which predicts an optical depth of roughly unity for 500 GeV (10 TeV) 
gamma rays from a redshift of 0.2 (0.05).

\subsection{Bright X-ray Flares}

One of the highlights of the Swift mission has been the detection of bright X-ray flares in the afterglow phase of many 
GRBs~\cite{burrows05}. These flares can sometimes be as bright as the prompt component of the GRB itself. While the exact 
nature of the flares is not clear, with various different proposals for what may be causing them 
(e.g. \cite{guetta07,kobayashi07,krimm07}), there are reasonable expectations that these flares will result in the 
emission of delayed very high energy photons~\cite{wang06}. The first survey of X-ray flares from GRBs observed by Swift 
in the first year of operations~\cite{chincarini} shows that approximately one third of GRBs show significant flaring activity. 
Table~\ref{flare_table} lists the 10 flares from the flare catalog of ~\cite{chincarini,falcone} which were in the field of 
view of Milagro. 

\section{Data Analysis and Results}

A search for an excess of events above those due to the background was made for each of the 57 bursts
in Table~\ref{grb_table}, as well as for the 10 flares listed in Table~\ref{flare_table}. 
The number of events falling within a 1.6 degree bin was summed for the relevant duration. An 
estimate of the number of background events was made by characterizing the angular distribution of the background 
using two hours of data surrounding the burst (or flare), as described in \cite{atkins03b}. 
No significant emission was detected. The upper limits 
are given in the final column of Table~\ref{grb_table} 
and Table~\ref{flare_table}. For those bursts with known redshift, we compute 
the effect of the absorption, according to the model of \cite{primack05} and print the upper limits in bold. 

Among the most interesting limits presented here is the one for GRB 070521, recently detected by Swift. 
Assuming a redshift of 0.55 for the possible host~\cite{hattori}, the Milagro upper limit~\cite{saz} on the 
fluence is lower than the $\sim$1.8$\times10^{-5}$ erg cm$^{-2}$ measured by the Konus-Wind 
experiment~\cite{golenetskii} in the 20 keV -- 1 MeV energy range. It should be noted, however, that 
\cite{atteia} have calculated a ``pseudo-redshift'' for this burst of pz= 2.28$\pm$0.45, and on these 
grounds, claim it is incompatible with the spectroscopic redshift of 0.55.

\section{Acknowledgments}

We have used GCN Notices to select raw data for archiving and use in this search, and
we are grateful for the hard work of the GCN team, especially Scott Barthelmy. We are 
grateful to Kevin Hurley for his help with the IPN bursts. We acknowledge 
Scott Delay and Michael Schneider for their dedicated efforts in the construction and 
maintenance of the Milagro experiment.  This work has been supported by the National Science 
Foundation (under grants 
PHY-0245234, 
-0302000, 
-0400424, 
-0504201, 
-0601080, 
and
ATM-0002744) 
the US Department of Energy (Office of High-Energy Physics and 
Office of Nuclear Physics), Los Alamos National Laboratory, the University of
California, and the Institute of Geophysics and Planetary Physics.




\def \atel {The Astronomer's Telegram}
\def \apj {ApJ}
\def \aj {AJ}
\def \apjl {ApJL}
\def \mnras {MNRAS}
\def \iaucirc {IAUCIRC}
\def \em { }
\def \aap {A\&A}
\def \nat {Nature}
\def \araa {Anual Review of Astronomy and Astrophysics}

\nocite{ref4}
\nocite{ref5}
\nocite{ref6}
\nocite{ref7}

\newpage
\begin{table}
\begin{minipage}[t]{2.05\columnwidth}%
\begin{small}
\begin{tabular}{llllll}
\hline
GRB &
Dur. & 
$\theta$ & 
z & 
Inst. & 
UL \\
\hline
041219a & 520  	& 27 	& ... 	& INT. 		& 5.8e-6 \\
050124	& 4  	& 23	& ...	& Swift		& 3.0e-7 \\
050213 	& 17  	& 23 	& ...	& IPN 		& 1.3e-6 \\
050319  & 15   	& 45    & 3.24 	& Swift	  	& ... \\
050402  & 8   	& 40    & ... 	& Swift	  	& 2.1e-6 \\
050412  & 26   	& 37    & ... 	& Swift	  	& 1.7e-6 \\
050502  & 20   	& 43    & 3.793 & INT. 		& ...  \\
050504  & 80   	& 28    & ... 	& INT.  	& 1.3e-6 \\
050505  & 60   	& 29    & 4.3 	& Swift		& ...  \\
050509b & 0.128 & 10    & 0.226?& Swift	 	& \textbf{1.1e-6} \\
050522  & 15   	& 23    & ... 	& INT.		& 5.1e-7  \\
050607  & 26.5 	& 29    & ... 	& Swift	 	& 8.9e-7  \\
050703 	& 26	& 26  	&...	& IPN 		& 1.2e-6  \\
050712  & 35   	& 39    & ... 	& Swift	   	& 2.5e-6  \\
050713b & 30   	& 44    & ... 	& Swift	   	& 4.0e-6  \\
050715  & 52   	& 37    & ... 	& Swift	   	& 1.7e-6  \\
050716  & 69   	& 30    & ... 	& Swift	   	& 1.6e-6  \\
050820  & 20   	& 22    & 2.612 & Swift	  	& ...  \\
051103  & 0.17 	& 50 	& 0.001?& IPN		& \textbf{4.2e-6} \\ 
051109  & 36   	& 9.7   & 2.346 & Swift	  	& ... \\
051111  & 20   	& 44  & 1.55 	& Swift		& ...  \\
051211b & 80	& 33	& ...	& INT.		& 2.6e-6	\\
051221	& 1.4	& 42	& 0.55	& Swift		& \textbf{9.8e-4} \\
051221b	& 61	& 26	& ... 	& Swift		& 1.8e-6 \\
060102	& 20	& 40	& ... 	& Swift		& 2.0e-6 \\
060109	& 10	& 22	& ... 	& Swift		& 4.1e-7 \\
060110	& 15	& 43	& ... 	& Swift		& 3.0e-6 \\
060111b	& 59	& 37	& ... 	& Swift		& 2.3e-6 \\
060114	& 100	& 41	& ... 	& INT.		& 5.1e-6 \\
060204b	& 134	& 31	& ... 	& Swift		& 2.7e-6 \\
060210	& 5	& 43	& 3.91 	& Swift		& ... \\
060218	& 10	& 45	& 0.03 	& Swift			& \textbf{3.8e-5} \\
060306	& 30	& 46	& ... 	& Swift		& 7.2e-6 \\
060312	& 30	& 44	& ... 	& Swift		& 3.3e-6 \\
060313	& 0.7	& 47	& ... 	& Swift		& 2.7e-6 \\
060403	& 25	& 28	& ... 	& Swift		& 1.0e-6 \\
060427b	& 0.22	& 16	& ... 	& IPN		& 2.1e-7 \\
060428b	& 58	& 27	& ... 	& Swift		& 1.1e-6 \\
060507	& 185	& 47	& ... 	& Swift		& 1.8e-5 \\
060510b	& 330	& 43	& 4.9 	& Swift		& ... \\
060515	& 52	& 42	& ... 	& Swift		& 9.6e-6 \\
060712	& 26	& 35	& ... 	& Swift		& 3.8e-6 \\
060814	& 146	& 23	& ... 	& Swift		& 2.5e-6 \\
060904A	& 80	& 14	& ... 	& Swift		& 2.4e-6 \\
060906	& 43.6	& 29	& 3.685 & Swift		& ... \\
061002	& 20	& 45	& ... 	& Swift		& 4.0e-6 \\
061126	& 191	& 28	& ... 	& Swift		& 4.3e-6 \\
061210	& 0.8	& 23	& 0.41? & Swift		& \textbf{6.1e-6} \\
061222a	& 115	& 30.	& ... 	& Swift		& 5.6e-6 \\
070103	& 19	& 39	& ...	& Swift		& 1.3e-6\\
070125	& 60	& 9.5	& 1.547	& IPN		& \textbf{4.5e-4} \\
070129	& 460	& 31	& ...	& Swift		& 1.9e-6 \\
070208	& 50	& 32	& 1.165	& Swift		& \textbf{4.0e-4} \\
070311	& 50	& 33	& ...	& INT.		& 2.0e-6 \\
070402	& 12	& 12	& ...	& IPN		& 4.6e-7 \\
070521	& 60	& 8.7	& 0.55?	& Swift		& \textbf{1.1e-5} \\
070529	& 120	& 45	& 2.5	& Swift		& ... \\

\hline
\end{tabular}
\end{small}
\caption{Recent GRBs in the Milagro field of view. Column 1 is the GRB name. Column 2 gives the 
duration of the burst (in seconds), column 3 the zenith angle for Milagro (in degrees), column 4 
the measured redshift, when it exists, column 5 the satellite(s) detecting the GRB, and column 6 
gives the Milagro 99\% confidence upper limit on the 0.2--20 TeV fluence in erg cm$^{-2}$. Numbers 
in bold take into account absorption by the EBL (using the Primack 05 
model) for a redshift given in column 4. Those with three dots in column 6 imply the redshifts are 
so high that all the emission is expected to be absorbed.\label{grb_table}}
\end{minipage}
\end{table}

\begin{table}
\begin{minipage}[l]{0.95\columnwidth}%
\begin{small}
\begin{tabular}{llllll}
\hline
GRB &
UTC & 
$\mathrm{T_i}$& 
$\mathrm{T_f}$& 
$\theta$ & 
UL \\
\hline
050607  &33082.8&  94	&  255	&  29	&  2.7e-6\\
050607	&33082.8& 255	&  640	&  29	&  3.2e-6\\
050712 	&50427.5&  88	&  564	&  38	&  8.8e-6\\
050712  &50427.5& 302	&  435	&  38	&  4.1e-6\\
050712  &50427.5& 415 	&  590	&  30.	&  2.9e-6\\
050712  &50427.5& 788 	&  952	&  37	&  2.3e-6\\
050716  &45363.6& 155 	&  211	&  31	&  1.2e-6\\
050716  &45363.6& 315 	&  483	&  32	&  1.8e-6\\
050820	&23693.1& 200	&  382	&  21	& ... \\
060109	&60881.2& 4305	&  6740 &  6.6	&  4.8e-6\\

\hline
\end{tabular}
\end{small}
\caption{List of X-ray Flares. Column 1 gives the Swift GRB name during which they occurred. Column 2 
gives the trigger time in UTC second of the day. Columns 3 and 4 give the start and end times of 
the flare relative to the trigger time given in Column 2. Column 5 gives the zenith angle for 
Milagro. Column 6 gives the Milagro 99\% upper limit on the 0.2--20 TeV fluence in 
erg cm$^{-2}$.\label{flare_table} As in Table~\ref{grb_table}, three dots implies the redshift is 
so high that all emission is expected to be absorbed.}
\end{minipage}
\end{table}

\end{document}